\documentclass[letterpaper, 10pt, conference]{IEEEtran}
\usepackage{latexsym}
\usepackage{ascmac}
\usepackage{graphicx}
\usepackage{wrapfig}
\usepackage[cmex10]{amsmath}
\usepackage{amssymb, amsfonts, amsthm,longtable}
\usepackage{array}
\usepackage{algorithm}
\floatstyle{plaintop}
\restylefloat{algorithm}
\usepackage{algorithmic}
\usepackage{url}
\usepackage{flushend}
\usepackage{fixltx2e}

\newtheorem{thm}{Theorem}
\newtheorem{lem}[thm]{Lemma}
\newtheorem{cor}[thm]{Corollary}
\newtheorem{defi}[thm]{Definition}

\begin{document}

\title{Average Shortest Path Length of Graphs of Diameter 3}
\author{\IEEEauthorblockN{Nobutaka Shimizu}
  \IEEEauthorblockA{Dept.\ of Mathematical Informatics,\\Graduate School of Information Science and Technology,\\The University of Tokyo\\
    Email: nobutaka\_shimizu@mist.i.u-tokyo.ac.jp}
  \and
  \IEEEauthorblockN{Ryuhei Mori}
  \IEEEauthorblockA{Dept.\ of Mathematical and Computing Sciences,\\
    School of Computing,\\
    Tokyo Institute of Technology\\
    Email: mori@c.titech.ac.jp}}

\maketitle
\begin{abstract}
  A network topology with low {\em average shortest path length (ASPL)} provides efficient data transmission while the number of nodes and the number of links incident to each node are often limited due to physical constraints. In this paper, we consider the construction of low ASPL graphs under these constraints by using stochastic local search (SLS) algorithms.
  Since the ASPL cannot be calculated efficiently, the ASPL is not suitable for the evaluation function of SLS algorithms.
  We first derive an equality and bounds for the ASPL of graphs of diameter 3. On the basis of the simplest upper bound of the ASPL, we propose to use $3\triangle+2\Box$ as the evaluation function for graphs of diameter 3 where $\triangle$ and $\Box$ denote the number of triangles and squares in a graph, respectively.
  We show that the proposed evaluation function can be evaluated in $O(1)$ time as the number of nodes and the maximum degree tend to infinity by using some data tables. By using the simulated annealing with the proposed evaluation function, we construct low ASPL regular graphs of diameter 3 with 10\,000 nodes.
\end{abstract}


\IEEEpeerreviewmaketitle

\section{Introduction}
\par The network topology is significant on the performance of the interconnection network, and thus various kinds of topologies have been proposed and analyzed in many areas such as Datacenter network, High Performance Computing (HPC), Peer-to-peer system, Network-on-Chip (NoC), and so on \cite{deBruijn,Jellyfish,analysis,meshstar,3-Dnoc,datacenter,swapandrandom}. In these areas, topologies with small diameter and low average shortest path length (ASPL) are desired. 
For example, the Hypercube topologies and the de~Bruijn graphs are known to be effective for NoC architectures~\cite{3-Dnoc} and peer-to-peer networks~\cite{deBruijn, analysis}, respectively. Some topologies based on randomness have been proposed and explored in the area of Datacenter network and HPC~\cite{Jellyfish,datacenter,swapandrandom}.
\par The problem of finding the maximum graph for given maximum degree and diameter is called the {\em degree/diameter problem},  and has been studied in graph theory \cite{MooreBound}. 
However, in practice, the order (the number of nodes) is often limited due to several reasons.
Hence, for many applications, we are given order and maximum degrees, and try to minimize the diameter and the ASPL\@.
This problem is called the {\em order/degree problem}, and has not been studied sufficiently.
Recently, the importance of the order/degree problem is pointed out~\cite{opticalforHPC}.
Some of the authors of~\cite{opticalforHPC} and their coworkers held a competition called ``Graph Golf'' on the order/degree problem~\cite{GraphGolfWeb}.

The stochastic local search (SLS) is a framework of approximation algorithms for general optimization problems.
In this paper, we consider the construction of low ASPL graphs of given order and given maximum degrees by using SLS algorithms.
In a SLS algorithm, 
an initial feasible solution is generated (possibly be a random solution). Then, the solution is iteratively replaced by one of its neighborhoods. 
For our problem, we use a simple local modification procedure called {\em switch} for defining the neighborhoods.
In this paper, we assume that SLS algorithms use the {\em evaluation function} which represents a quality of feasible solutions~\cite{SLSbook}.
Since huge number of feasible solutions must be evaluated in SLS algorithms in general, efficient evaluation algorithms are strongly desired. 
However, the ASPL cannot be calculated efficiently at least in our knowledge, and hence is not suitable for the evaluation function.

In this paper, we assume that for given $n$ and $d$, the $d$-regular random graph of order $n$ has the diameter 3 with high probability, and propose the evaluation function for graphs of diameter 3 which can be evaluated efficiently.
More precisely, first, we derive an equality and bounds for the ASPL of graphs of diameter 3.
On the basis of the simplest upper bound of the ASPL, we propose to use $3\triangle+2\Box$ as the evaluation function of SLS algorithms where $\triangle$ and $\Box$ denote the number of triangles and squares in a graph, respectively.
The proposed evaluation function can be evaluated in $O(1)$ time as $n$ and $d$ tend to infinity by using some data tables. By using the proposed evaluation function in the iterative first improvement (IFI) and the simulated annealing (SA), we construct low ASPL regular graphs of diameter 3 with 10\,000 nodes, which are the best graphs in the competition Graph Golf.

\par The remainder of this paper is organized as follows. In Section~\ref{preliminarysec}, notions and notations used in this paper are introduced. In Section~\ref{theoremsec}, the equality and bounds for the ASPL of graphs of diameter 3 are shown.
In Section~\ref{algorithmsec}, we propose the new evaluation function, and show an efficient algorithm for calculating the new evaluation function using data tables. In Section~\ref{experimentsec}, we show results of numerical experiments. Section~\ref{concludesec} concludes this paper.

\section{Preliminaries} \label{preliminarysec}
\subsection{Notations and definitions} \label{Notations and Definitions}
A graph $G$ is a pair of two finite sets $V$ and $E$ where every element of $E$ is a subset of $V$ of size 2.
Each element of $V$ is called a {\em node}. Each element of $E$ is called an {\em edge}.
Let $e_{ij}:=\{i,j\}$ for $\{i,j\}\in E$.
The number $n$ of nodes is called the {\em order of $G$}.
For nodes $u,v \in V$, $u$ is said to be {\em connected to $v$} if there is an edge $\{u,v\}\in E$. The {\em degree} $d_v$ of a node $v$ is the number of nodes connected to $v$.
 A graph is {\em $d$-regular} if the degrees of all the nodes are equal to $d$.
\par For two graphs $G=(V,E)$ and $G'=(V',E')$, {\em $G'$ is a subgraph of $G$} if $V' \subseteq V$ and $E' \subseteq E$. If $G'$ is a subgraph of $G$, we say $G$ {\em contains} $G'$.
We say {\em $G$ is isomorphic to $G'$} if there is a bijection $f : V \rightarrow V'$ that satisfies the following condition:
\[
\forall \{u,v\} \subseteq V,\hspace{2em} \{u,v\} \in E \Longleftrightarrow \{f(u),f(v)\} \in E'.
\]
For two graphs $G$ and $H$, {\em the number of $H$ in $G$} is the number of subgraphs of $G$ that are isomorphic to $H$.
\par For two nodes $s$ and $t$, a {\em $s$-$t$ path} or a {\em path} is a graph $P=(V,E)$ where
\begin{align*}
  V = \{s=v_0,v_1,\ldots,v_\ell=t\}, \qquad E=\bigcup_{k=0}^{\ell-1} \{\{v_k,v_{k+1}\}\}
\end{align*}
for $v_0,v_1,\ldots,v_\ell$ which are distinct nodes. Here, $\ell$ is called the {\em length} of $s$-$t$ path $P$. If $P=(V,E)$ is a $s$-$t$ path of length $\ell \geq 2$, a graph $C=(V,E\cup\{\{s,t\}\})$ is called a {\em cycle}. The length of the cycle $C$ is $\ell+1$.
We call a cycle of length 3 and 4 a {\em triangle} and a {\em square}, respectively.

\begin{defi}
  For $k=0,1,\ldots$ a {\em $k$-multiple triangle} is a graph $\mathrm{Tri}^{(k)}=(V^{(k)},E^{(k)})$ where
  \begin{align*}
    V^{(k)} &:= \begin{cases}
      \{i,j\}, &\hspace{2em} k=0 \\
      V^{(k-1)}\cup \{v_k\}, &\hspace{2em} k>0
    \end{cases} \\
    E^{(k)} &:= \begin{cases}
      \{\{i,j\}\}, &\hspace{2em} k=0 \\
      E^{(k-1)} \cup \{\{i,v_{k}\},\{v_{k},j\}\}, &\hspace{2em} k>0.
    \end{cases}
  \end{align*}
  Such a $k$-multiple triangle is called a {\em $k$-multiple triangle sharing $i,j$} when we specify the two nodes $i$ and $j$.
\end{defi}
\begin{defi}
  For $k=0,1,\ldots$ a {\em $k$-multiple square} is a graph $\mathrm{Squ}^{(k)}=(V^{(k)},E^{(k)})$ where
  \begin{align*}
    V^{(k)} &:= \begin{cases}
      \{i,j,v_0\}, &\hspace{2em} k=0 \\
      V^{(k-1)}\cup \{v_k\}, &\hspace{2em} k>0
    \end{cases} \\
    E^{(k)} &:= \begin{cases}
      \{\{i,v_0\},\{v_0,j\}\}, &\hspace{2em} k=0 \\
      E^{(k-1)} \cup \{\{i,v_{k}\},\{v_{k},j\}\}, &\hspace{2em} k>0.
    \end{cases}
  \end{align*}
  Such a $k$-multiple square is called a {$k$-multiple square sharing $i,j$} when we specify the two nodes $i$ and $j$.
\end{defi}
\begin{figure}
  \centering
  \includegraphics[width=7cm]{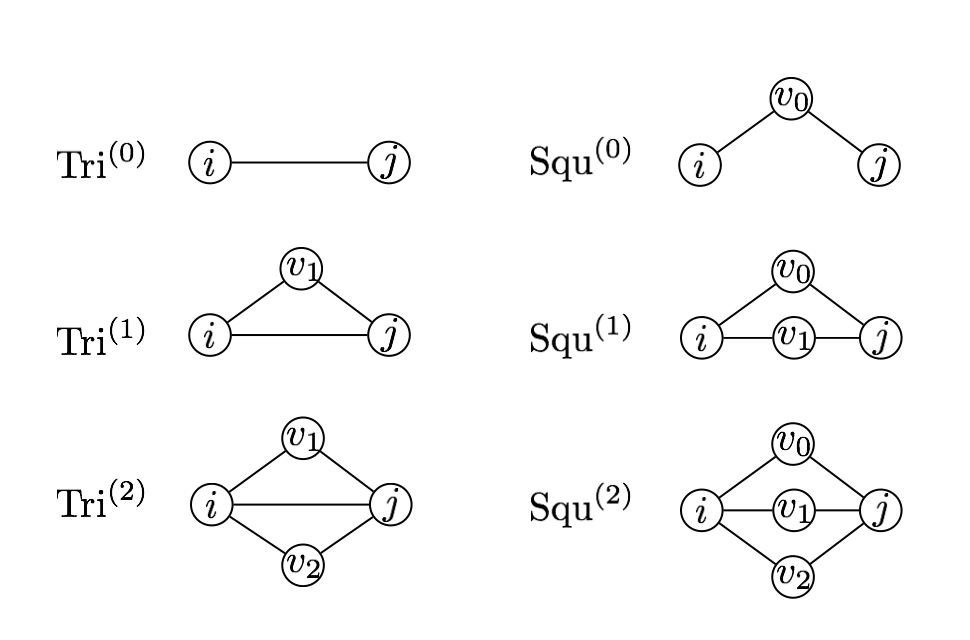}
  \caption{Some examples of $k$-multiple triangles and $k$-multiple squares.\label{figures}}
\end{figure}
Fig.~\ref{figures} shows $k$-multiple triangles and $k$-multiple squares sharing $i,j$ for $k=0,1,2$.
Note that the $1$-multiple triangle and the $1$-multiple square are equivalent to the triangle and the square, respectively.
The number of $k$-multiple triangles and the number of $k$-multiple squares in a graph are denoted by $\triangle^{(k)}$ and $\Box^{(k)}$, respectively.
Let $\triangle := \triangle^{(1)}$ and $\Box:=\Box^{(1)}$.
\par For a graph $G=(V,E)$ and $s,t \in V$, a $s$-$t$ path whose length is minimum among all $s$-$t$ paths contained in $G$ is called the {\em shortest path between $s$ and $t$}.
The length of the shortest path between $s$ and $t$ is called the {\em distance between $s$ and $t$}, denoted by $\mathrm{dist}(s,t)$. If $G$ contains no $s$-$t$ paths, we define $\mathrm{dist}(s, t) = \infty$. The {\em diameter} of $G$ is defined by $\max_{\{i,j\}\subseteq V, i \neq j} \mathrm{dist}(i,j)$.

\begin{defi} \label{ASPLdefi}
  For a graph $G=(V,E)$, {\em the average shortest path length of $G$} is defined by
  \[
  \mathrm{ASPL}(G) := \frac{2}{n(n-1)}\sum_{\{i,j\}\subseteq V,\, i\ne j} \mathrm{dist}(i,j).
  \]
\end{defi}
In this paper, we consider the following graph optimization problem: For given $n$ and $d$,
\begin{displaymath}
  \begin{array}{ll}
    \mbox{{\rm minimize:} } & \mathrm{ASPL}(G) \\
    \mbox{{\rm subject to:} }& |V(G)|=n \\
    &d_v = d, \hspace{2em} \forall v \in V(G). \\
  \end{array}
\end{displaymath}
Here, $V(G)$ denotes the node set of $G$.
Note that the set of feasible solutions of this problem is the set of all $d$-regular graphs of order $n$.
\begin{figure}
  \centering
    \includegraphics[width=6.25cm]{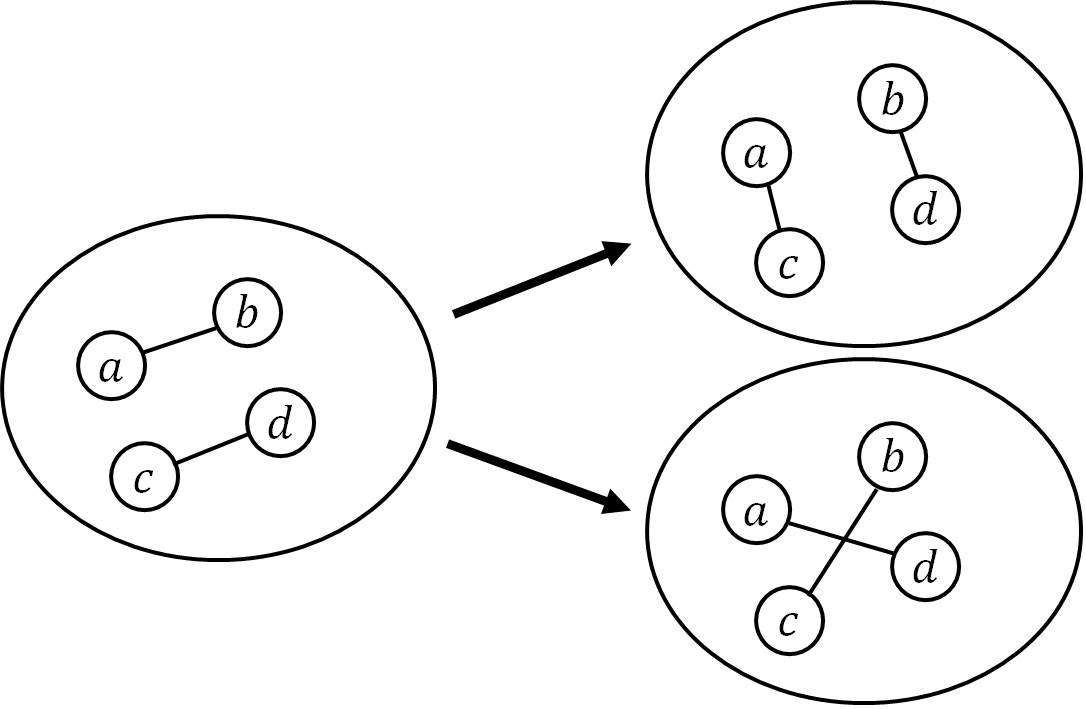}
  \caption{An image of switch for $e_{ab}$ and $e_{cd}$. \label{switch}}
\end{figure}
\subsection{Stochastic local search for our problem}
\par In this paper, we consider SLS algorithms described below.
First, an initial $d$-regular graph $G$ of order $n$ is chosen randomly.
Then, $G$ is iteratively replaced by a local modification procedure called {\em switch}.
In the switch procedure, a pair of edges $(e_{ab}, e_{cd})$ is chosen and replaced by $(e_{ac}, e_{bd})$ or $(e_{ad}, e_{bc})$ as in Fig.~\ref{switch}.
If the new graph $G'$ is not $d$-regular, i.e., at least one of the edges introduced by the switch already exists in $G$, a new pair of edges in $G$ is chosen for the switch until a $d$-regular graph $G'$ is found.
Here, the $d$-regular graph $G'$, which can be obtained by a single switch of $G$, is called {\em a neighborhood} of $G$.
Once a neighborhood $G'$ is found, $G'$ is evaluated by {\em the evaluation function}. According to the value of the evaluation function on $G'$, it is determined whether or not the current graph $G$ is replaced by $G'$.
This updating procedure continues until further improvement is not expected.

It is natural to define the evaluation function by the ASPL of $G'$. In order to calculate the ASPL of a $d$-regular graph of order $n$, the distances of all node pairs have to be calculated,
which takes $O(n^2d)$ time by using the breadth first search. Since it is not sufficiently efficient, 
low ASPL graphs cannot be obtained within a reasonable time for large $n$ and $d$.

In this paper, we consider an optimization of $d$-regular graphs of diameter 3.
The Moore bound implies that any graph of diameter 2 satisfies $n \leq d^2+1$ and any graph of diameter 3 satisfies $n \leq d^3-d^2+d+1$ \cite{MooreBound}.
Furthermore, almost all graphs of diameter 2 satisfy $n<d^2$ \cite{MooreBound,erd}. Therefore, if $n\ge d^2$ and $n$ is close to $d^2$, then a random $d$-regular graph of order $n$ is expected to have diameter 3 with high probability.
For such $n$ and $d$,
if we choose a random $d$-regular graph of order $n$ as the initial graph of a SLS algorithm, the diameter of every graph appearing in the SLS algorithm is empirically 3.
By assuming that the diameter of graphs found by the SLS algorithm is always 3, the ASPL can be calculated in $O(d)$ time\footnote{The ASPL of graph of diameter 3 is determined by the adjacency matrix $A$ of the graph and its square $A^2$.
Therefore, we can calculate the difference of the ASPLs of $G$ and $G'$ by updating $A$ and $A^2$. The updating requires $O(d)$ time since $O(d)$ elements of $A^2$ change.},
which is still not sufficiently efficient for large $n$.
In this paper, we give an upper bound of the ASPL of graphs of diameter 3, and propose SLS algorithms which uses the upper bound as the evaluation function. It will be shown that the upper bound can be calculated in $O(1)$ time.

\section{Bounds for the ASPL of graphs of diameter 3}\label{theoremsec}
In this section, an equality and bounds for the ASPL of graphs of diameter 3 are shown.
They are stated not only for $d$-regular graphs but also for general graphs.
\begin{defi}
  For a graph $G=(V,E)$, let $V_2 := \{\{i,j\}\,:\, i\in V, j\in V, i\ne j\}$.
  For $\{i,j\} \in V_2$,
  \[
  W_{ij}:=(V \backslash \{i,j\}) \cup \{0\}.
  \]
  For $\{i,j\} \in V_2$ and $k \in W_{ij}$,
  \[
  S^{i,j}_k :=\begin{cases}
  \{\{i,j\} \,:\, \{i,j\}\in E \}, &\hspace{2em} k=0 \\
  \{\{i,j\} \,:\, \{i,k\},\{k,j\} \in E \}, &\hspace{2em} k\ne 0.
  \end{cases}
  \]
  For $m=1,2,\ldots,n-1$,
  \begin{equation}\label{defiT}
    T(m) := \sum_{\{i,j\} \in V_2} \sum_{\substack{K \subseteq W_{ij}\\ |K|=m}} \left| \bigcap_{k \in K} S^{i,j}_k \right|.
  \end{equation}
\end{defi}
Note that $S^{i,j}_k$ is either $\{\{i,j\}\}$ or the empty set.
\begin{lem}\label{lemmatwo}
  For a graph $G=(V,E)$,
  \[
  n_1+n_2 = \sum_{m=1}^{n-1} (-1)^{m-1} T(m).
  \]
  where $n_k = \#\{\{i,j\} \in V_2 : \mathrm{dist}(i,j) = k \}$ for $k=1,2$.
\end{lem}
\begin{IEEEproof}
  For $\{i, j\}\in V_2$, it holds
  \[
  \{i,j\}\in \bigcup_{k \in W_{ij}} S^{i,j}_k \, \Longleftrightarrow \, \mathrm{dist}(i,j) \leq 2.
  \]
  We can obtain the lemma by using the inclusion-exclusion principle.
  \begin{eqnarray}
    n_1+n_2
    &=& \sum_{\{i,j\} \in V_2} \left| \bigcup_{k\in W_{ij}} S^{i,j}_k \right| \nonumber\\
    &=& \sum_{\{i,j\} \in V_2} \sum_{m=1}^{n-1} (-1)^{m-1}  \sum_{\substack{K \subseteq W_{ij}\\ |K|=m}} \left| \bigcap_{k \in K} S^{i,j}_k \right| \nonumber\\
    &=& \sum_{m=1}^{n-1} (-1)^{m-1}T(m). \label{eq:incexc}
  \end{eqnarray}
\end{IEEEproof}

\begin{lem}\label{lemmaone}
  \begin{equation}
    T(m) = \begin{cases}
      \frac{1}{2} \sum_{k \in V} d_k^2, &\hspace{2em} m=1 \\
      3\triangle+2\Box, &\hspace{2em} m=2 \\
      \triangle^{(m-1)}+\Box^{(m-1)}, &\hspace{2em} m \geq 3.
    \end{cases}
  \end{equation}
\end{lem}

\begin{IEEEproof}
  When $m=1$
  \begin{eqnarray*}
    T(1) &=& \sum_{\{i,j\} \in V_2} \left| S^{i,j}_0 \right| + \sum_{\substack{\{i,j\}\in V_2\\k\in W_{ij}\backslash \{0\}}} \left|S^{i,j}_k\right| \\
    &=& |E| + \sum_{k \in V} \binom{d_k}{2}  
    = \frac{1}{2} \sum_{k \in V}d_k^2.
  \end{eqnarray*}
  Here, we used the fact $|E|=\sum_{k\in V} d_k/2$.
  Assume $m\ge 2$. For $K\subseteq W_{ij}$ and $|K|=m$, $|\bigcap_{k\in K} S^{i,j}_k|=1$ if
  \begin{equation} \label{lemmaone1}
    \{i,j\}\in \bigcap_{k\in K} S^{i,j}_{k}
  \end{equation}
  and $|\bigcap_{k\in K} S^{i,j}_k|=0$ otherwise.
  If $0\in K$,~\eqref{lemmaone1} holds if and only if there is an $(m-1)$-multiple triangle sharing $i,j$ whose node set is $\{i,j\}\cup K\setminus\{0\}$.
  Similarly, if $0\notin K$, the equality~\eqref{lemmaone1} holds if and only if there is an $(m-1)$-multiple square sharing $i,j$ whose node set is $\{i,j\}\cup K$.
  \par Assume $m=2$. If there is a triangle consisting of three nodes $i$, $j$ and $k$ and of edge set $\{e_{ij},e_{jk},e_{ki}\}$, it holds
  \begin{align*}
     \{i,j\}&\in S^{i,j}_0 \cap S^{i,j}_k,&
     \{j,k\}&\in S^{j,k}_0 \cap S^{j,k}_i,\\
     \{k,i\}&\in S^{k,i}_0 \cap S^{k,i}_j.
  \end{align*}
  If there is a square consisting of four nodes $a$, $b$, $c$ and $d$ and of edge set $\{e_{ab},e_{bc},e_{cd},e_{da}\}$, it holds
  \begin{align*}
     \{a,c\}&\in S^{a,c}_b \cap S^{a,c}_d,&
     \{b,d\}&\in S^{b,d}_a \cap S^{b,d}_c.
  \end{align*}
  Therefore,
  \begin{align*}
    T(2)&= \sum_{\substack{\{i,j\} \in V_2\\k,l\in W_{ij}}} \left|S^{i,j}_k \cap S^{i,j}_l \right| \\
    &= \sum_{\substack{\{i,j\} \in V_2\\k\in W_{ij}\backslash \{0\}}} \left|S^{i,j}_0 \cap S^{i,j}_k \right| 
    + \sum_{\substack{\{i,j\} \in V_2\\k,l \in W_{ij} \backslash \{0\}}} \left|S^{i,j}_k \cap S^{i,j}_l \right| \\
    &= 3\triangle + 2\Box.
  \end{align*}
  \par Assume $m \geq 3$.
  If there is an $(m-1)$-multiple triangle sharing $i,j$ consisting of nodes $\{i,j\}\cup K$ for some $K\subseteq W_{ij}\setminus \{0\}$,
  it holds
     $\{i,j\}\in \bigcap_{k\in K\cup \{0\}} S^{i,j}_k$ and $\left|K \cup \{0\} \right| = m$.
Similarly, if there is an $(m-1)$-multiple square sharing $i,j$ consisting of nodes $\{i,j\}\cup K$ for some $K\subseteq W_{ij}\setminus \{0\}$,
     it holds $\{i,j\}\in \bigcap_{k\in K} S^{i,j}_k$ and $\left|K \right|=m$.
  Therefore, $T(m)=\triangle^{(m-1)}+\Box^{(m-1)}$.
\end{IEEEproof}

\begin{thm}
  For a graph $G=(V,E)$ of diameter 3,
  \begin{align*}
    \mathrm{ASPL}(G) &= 3-\frac{2}{n(n-1)}\left(|E|+\sum_{m=1}^{n-1} (-1)^{m-1} T(m) \right).
  \end{align*}
\end{thm}
\begin{IEEEproof}
  $\mathrm{ASPL}(G)$ can be represented by
  \[
  \mathrm{ASPL}(G) = \frac{2}{n(n-1)}(n_1+2n_2+3n_3)
  \]
  where $n_k = \#\{\{i,j\} \in V_2 : \mathrm{dist}(i,j) = k \}$ for $k=1,2,3$.
  From Lemma \ref{lemmatwo} and the fact of $n_1=|E|$,
  \begin{align*}
    n_1+2n_2+3n_3 
    &= 3(n_1+n_2+n_3)-n_1-(n_1+n_2) \\
    &= \frac{3n(n-1)}{2}-|E|-\sum_{m=1}^{n-1}(-1)^{m-1}T(m).
  \end{align*}
\end{IEEEproof}
\begin{thm}\label{thm1}
  When $t$ is even
  \begin{align} \label{ASPLlowerbound}
    \mathrm{ASPL}(G) \leq 3-\frac{2}{n(n-1)}\left(|E|+\sum_{m=1}^{t} (-1)^{m-1} T(m) \right).
  \end{align}
  When $t$ is odd
  \begin{align} \label{ASPLupperbound}
    \mathrm{ASPL}(G) \geq 3-\frac{2}{n(n-1)}\left(|E|+\sum_{m=1}^{t} (-1)^{m-1} T(m) \right).
  \end{align}
\end{thm}
\begin{IEEEproof}
  When $t$ is even, by applying the Bonferroni inequality to~\eqref{eq:incexc}~\cite{Bon}, one obtains
  \[
  n_1+n_2 \geq \sum_{m=1}^t (-1)^{m-1}T(m).
  \]
  The theorem for odd $t$ can be proved in the same way.
\end{IEEEproof}
By substituting $t=1,2$ into~\eqref{ASPLlowerbound} and~\eqref{ASPLupperbound}, respectively, the following corollaries are obtained.
\begin{cor}
  For a $d$-regular graph $G$ of order $n$ and diameter 3,
  \begin{equation} \label{Moore}
    \mathrm{ASPL}(G) \geq 3-\frac{d(d+1)}{n-1}.
  \end{equation}
\end{cor}
\begin{cor}
  For a $d$-regular graph $G$ of order $n$ and diameter 3,
  \begin{equation} \label{ASPLupper}
    \mathrm{ASPL}(G) \leq 3 - \frac{d(d+1)}{n-1} + \frac{6\triangle+4\Box}{n(n-1)}.
  \end{equation}
\end{cor}
\par The lower bound (\ref{Moore}) agrees with the {\em Moore bound} \cite{MooreBound}.

\section{Proposed evaluation function and algorithm}\label{algorithmsec}
\subsection{SLS using the upper bound as the evaluation function}
In this paper, we propose SLS using~\eqref{ASPLupper} as the evaluation function.
More precisely, the evaluation function $g$ in our SLS is defined by
\begin{equation} \label{evalfunc}
  g(G) = 3\triangle+2\Box.
\end{equation}
Since the minimization of~\eqref{evalfunc} is equivalent to the minimization of~\eqref{ASPLupper}, we can expect that the SLS algorithm using
the above evaluation function finds a graph with low ASPL if all graphs appearing in the SLS algorithm have diameter 3.
In this section, we show that a difference $g(G')-g(G)$ for a $d$-regular graph $G$ of order $n$ and its neighborhood $G'$ can be calculated in $O(1)$ time by using three two-dimensional arrays.
We also show that these three arrays can be updated in $O(d^2)$ time.
In SLS algorithms, the frequency of the evaluations is much higher than the frequency of the updates.
Therefore, our SLS is much faster than SLS using the ASPL as the evaluation function, which needs $O(d)$ time both for the evaluation and the update.

For the current graph $G$, the three two-dimensional arrays used for the $O(1)$-time evaluation are defined by
\begin{align*}
  {\rm T1}[i][j] &:= \begin{cases}
    1, &\hspace{2em} \{i,j\}\in E(G) \\
    0, &\hspace{2em} \{i,j\} \not \in E(G), \end{cases}\\
  {\rm T2}[i][j] &:= \#\{ \text{$i$-$j$ paths of length 2 in $G$}\}, \\
  {\rm T3}[i][j] &:= \#\{ \text{$i$-$j$ paths of length 3 in $G$}\}.
\end{align*}
Here, $E(G)$ is the edge set of $G$.
The construction of these three arrays takes $O(nd^3)$ time since the number of paths of length at most 3 in $G$ is $O(nd^3)$, and since $d=\Omega(n^{1/3})$ for guaranteeing the diameter of $G$ is equal to 3.
We first construct T1, T2 and T3 for the initial graph in the SLS.
Then, at each step of the SLS, we evaluate a neighborhood in $O(1)$ time by using T1, T2 and T3.
If the current graph is replaced, T1, T2 and T3 are updated in $O(d^2)$ time.
In the following, we show algorithms for the $O(1)$-time evaluation and the $O(d^2)$-time update.

\subsection{$O(1)$-time evaluation with arrays}
For the neighborhood $G'$, let $\triangle'$ be the number of triangles in $G'$ and $\Box'$ be the number of squares in $G'$.
Then, it holds
\begin{equation}
  g(G')-g(G) = 3(\triangle'-\triangle)+2(\Box'-\Box).
\end{equation}
We assume that a neighborhood $G'$ of $G$ is obtained by replacing $e_{ab},e_{cd} \in E(G)$ with $e_{ac},e_{bd} \not \in E(G)$.
The switch can be divided into two steps: the removal of $e_{ab},e_{cd}$ and the addition of $e_{ac},e_{bd}$.
First, we consider the removal of $e_{ab}$ and $e_{cd}$. Since no triangles contain both of $e_{ab}$ and $e_{cd}$, the number of triangles decreases by
\begin{equation}\label{dectri}
  {\rm T2}[a][b]+{\rm T2}[c][d].
\end{equation}
\par Next, we consider the addition of $e_{ac}$ and $e_{bd}$ to the graph obtained by the removal of $e_{ab}$ and $e_{cd}$.
Note that the arrays T1, T2 and T3 remain unchanged at this time.
After the removal of $e_{ab}$ and $e_{cd}$, the addition of $e_{ac}$ increases the number of triangles by
\begin{equation}\label{inctri-ac}
  {\rm T2}[a][c] - {\rm T1}[a][d]-{\rm T1}[b][c].
\end{equation}
Here, the terms ${\rm T1}[a][d]$ and ${\rm T1}[b][c]$ take account of non-existent paths $a$-$d$-$c$ and $a$-$b$-$c$, respectively.
Similarly, the addition of $e_{bd}$ increases the number of triangles by
\begin{equation}\label{inctri-bd}
  {\rm T2}[b][d] - {\rm T1}[a][d]-{\rm T1}[b][c].
\end{equation}
Hence, from~\eqref{dectri}, \eqref{inctri-ac} and \eqref{inctri-bd}, one obtains
\begin{align*}
  \triangle' - \triangle = &-{\rm T2}[a][b]-{\rm T2}[c][d]+{\rm T2}[a][c]+{\rm T2}[b][d]\\
  &- 2 ({\rm T1}[a][d]+{\rm T1}[b][c]).
\end{align*}

Next, we consider $\Box'-\Box$.
Since the number of squares that contain both of $e_{ab}$ and $e_{cd}$ is ${\rm T1}[a][d]{\rm T1}[b][c]$ (Recall that $e_{ac}$ and $e_{bd}$ are not in $E(G)$),
 the removal of $e_{ab}$ and $e_{cd}$ decreases the number of squares by
\begin{equation}\label{decsqu}
  {\rm T3}[a][b]+{\rm T3}[c][d] - {\rm T1}[a][d]{\rm T1}[b][c].
\end{equation}
After the removal of $e_{ab}$ and $e_{cd}$, the addition of $e_{ac}$ increases the number of squares by
\begin{equation} \label{incsqu-ac}
  {\rm T3}[a][c] - {\rm T2}[a][d] - {\rm T2}[b][c].
\end{equation}
Here, the terms ${\rm T2}[a][d]$ and ${\rm T2}[b][c]$ take account of non-existent paths  $a$-$*$-$d$-$c$ and $a$-$b$-$*$-$c$, respectively, where $*$ represents an arbitrary node (A non-existent path $a$-$b$-$d$-$c$ is not counted since $e_{bd}\notin E(G)$).
Similarly, after the addition of $e_{ac}$, the addition of $e_{bd}$ increases the number of squares by
\begin{equation} \label{incsqu-bd}
  {\rm T3}[b][d] - {\rm T2}[a][d] - {\rm T2}[b][c] + {\rm T1}[a][d]{\rm T1}[b][c].
\end{equation}
Here, the term ${\rm T1}[a][d]{\rm T1}[b][c]$ takes account of the path $b$-$c$-$a$-$d$ in $G'$.
Hence, from~\eqref{decsqu}, \eqref{incsqu-ac} and \eqref{incsqu-bd}, one obtains
\begin{align*}
  \Box'-\Box = &-{\rm T3}[a][b]-{\rm T3}[c][d]+{\rm T3}[a][c]+{\rm T3}[b][d]\\
  &-2({\rm T2}[a][d]+{\rm T2}[b][c]-{\rm T1}[a][d]{\rm T1}[b][c]).
\end{align*}
We conclude that we can calculate $g(G')-g(G)$ in $O(1)$ time if T1, T2 and T3 for $G$ are given.

\subsection{Array update} \label{Table Update}
When $e_{ab}$ is removed, T2 can be updated by decrementing ${\rm T2}[t][s]$ and ${\rm T2}[s][t]$ for all $s$-$t$ paths of length 2 using $e_{ab}$. Since the number of these paths is $O(d)$, this calculation can be done in $O(d)$ time. Also T3 can be updated by decrementing ${\rm T3}[s][t]$ and ${\rm T3}[t][s]$ for all $s$-$t$ paths of length 3 using $e_{ab}$. The number of these paths is $O(d^2)$ and thus this calculation can be done in $O(d^2)$ time.
When edges are added, the arrays can be updated in $O(d^2)$ time in the same way.

\section{Numerical experiments}\label{experimentsec}
\subsection{Accuracy of approximations}
In this section, we show by numerical experiments that the upper bounds and lower bounds obtained by Theorem~\ref{thm1} are also good approximations for the ASPL\@.
Table~\ref{result1} shows the relative error $(\widetilde{\mathrm{ASPL}}(G)-\mathrm{ASPL}(G))/\mathrm{ASPL}(G)$ of the approximations where $\widetilde{\mathrm{ASPL}}(G)$ denotes the bounds obtained by Theorem~\ref{thm1} for $t=1,2,3$,
and where $G$ denotes a random $d$-regular graph of order $n$ where $(n,d)=(4096,60),(4096,64),(10\,000,60),(10\,000,64)$. The diameters of these graphs are 3.
\begin{table}
  \newcolumntype{T}{>{\raggedleft\arraybackslash}p{4em}}
  \caption{The relative ASPL errors of approximations. \label{result1}}
  \centering
  \begin{tabular}{|c|T|T|T|}
    \hline
    $(n,d)$ & \multicolumn{1}{c|}{$t=1$} & \multicolumn{1}{c|}{$t=2$} & \multicolumn{1}{c|}{$t=3$} \\
    \hline
    $(4096,60)$  & $-0.1206$ & $0.0355$ & $-0.0074$\\
    $(4096,64)$  & $-0.1544$ & $0.0523$ & $-0.0124$\\
    $(10\,000,60)$ & $-0.0209$ & $0.0024$ & $-0.0002$\\
    $(10\,000,64)$ & $-0.0270$ & $0.0036$ & $-0.0003$\\
    \hline
  \end{tabular}
\end{table}
Table~\ref{result1} implies that the approximations are more accurate as $t$ increases.
Furthermore, Table~\ref{result1} also implies that the approximations are more accurate for sparse graphs.
It can be easily understood since $\triangle,\Box,\triangle^{(2)},\Box^{(2)},\ldots$ are small when a graph is sparse. 

\subsection{Construction by Iterative First Improvement}
The Iterative First Improvement (IFI) algorithm is one of the simplest SLS algorithms. The IFI algorithm evaluates neighborhoods one after another. When the IFI finds the neighborhood whose value of the evaluation function is lower than that of the current solution, it replaces the current solution by the neighborhood and then starts to evaluate the next neighborhood. The evaluation and replacement continues until the IFI algorithm finds the local optimum. We consider the IFI algorithm with the $O(1)$-time evaluation function proposed in Section~\ref{algorithmsec}.
\par In order to improve the efficiency of the IFI algorithm, we modify the IFI algorithm as follows: At every 50 replacements, the set of edges are sorted by the number of triangles and squares that contain the edge.
More precisely, all edges are sorted in descending order of $3\triangle_e + 2\Box_e$ where $\triangle_e$ and $\Box_e$ denote the number of triangles and squares that contain the edge $e$, respectively.
After the sort, all pairs of edges are chosen in the order of $(e_1,e_2), (e_1,e_3), \dotsc, (e_1,e_{|E|}), (e_2,e_3), (e_2,e_4),\dotsc$.
This modification is useful for finding efficiently a neighborhood improving the value of the evaluation function.

Table~\ref{randomASPL} shows the ASPL, the Moore bound~\eqref{Moore}, denoted by $L$, and the ASPL gap $(\mathrm{ASPL}(G)-L)/L$ for random $d$-regular graph $G$ of order $n$.
From these random graphs, we applied the above modified IFI algorithm, and obtained local optimals.
Table~\ref{result2} shows the ASPL, the ASPL gap for the local optimals and the required time for the modified IFI\@.
These numerical experiments are performed on MacBook Pro, Intel Core i7 2.6GHz.
The modified IFI found a local optimal within two days.
The obtained local optimals have smaller ASPL gap than the original random graphs.
Especially, for $(n,d)=(10\,000,60), (10\,000,64)$, the ASPL gaps are significantly improved.

\begin{table}[t]
  \centering
  \caption{ASPL of Random Graphs \label{randomASPL}}
  \begin{tabular}{|c|c|c|c|}
    \hline
    $(n,d)$ & ASPL & Moore & ASPL gap \\
    \hline
    $(4096,60)$ & 2.3951 & 2.1062 & $13.72\times 10^{-2}$ \\
    $(4096,64)$ & 2.3464 & 1.9841 & $18.26\times 10^{-2}$ \\
    $(10\,000,60)$ & 2.6901 & 2.6340 & $21.31\times 10^{-3}$ \\
    $(10\,000,64)$ & 2.6557 & 2.5840 & $27.76\times 10^{-3}$ \\
    \hline
  \end{tabular}
\end{table}
\begin{table}[t]
  \centering
  \caption{Results of the modified IFI \label{result2}}
  \begin{tabular}{|c|c|c|c|} \hline
    $(n,d)$ & ASPL & ASPL gap & Time (minutes) \\  \hline
    $(4096,60)$ &  2.3055 & $9.461\times 10^{-2}$ & 341\\
    $(4096,64)$ &  2.2536 & $13.58\times 10^{-2}$ & 336\\
    $(10\,000,60)$ & 2.6521 & $6.886\times 10^{-3}$ & 2429\\
    $(10\,000,64)$ & 2.6120 & $10.85\times 10^{-3}$ & 2222\\
    \hline
  \end{tabular}
\end{table}

\subsection{Construction by Simulated Annealing}
The Simulated Annealing (SA) is one of the most successful SLS algorithms.
The SA in our numerical experiments works as follows.
At each step, a neighborhood is chosen uniformly at random and evaluated. If the value $E'$ of evaluation function for the neighborhood is lower than that $E$ for the current solution, then the current solution is replaced by the neighborhood.
Also, even if the neighborhood has the higher evaluation value than the current solution, i.e., $E'>E$, the current graph is also replaced by the neighborhood with probability $P$ where
\[
P = \exp \left\{-\frac{E'-E}{T}\right\}.
\]
Here, $T$ is a parameter called the {\em temperature} which decreases at each step of SA\@.
The annealing schedule, which determines the temperature at each step, is crucial to the SA\@.
In the numerical experiments, we set $T(k):=11/ \ln (k+1)$ where $k$ denotes the number of graphs which are evaluated until that time.
Table~\ref{saresult} shows the ASPL of graphs constructed by applying the SA 60 days, and then applying the modified IFI.
They are the best graphs obtained in the Graph Golf~\cite{GraphGolfWeb}.
\begin{table}[t]
  \centering
  \caption{Results of the SA \label{saresult}}
  \begin{tabular}{|c|l|l|}
    \hline
    \multicolumn{1}{|c|}{Graph} & \multicolumn{1}{c|}{ASPL} & \multicolumn{1}{c|}{ASPL gap} \\
    \hline
    $(10\,000,60)$ & 2.6502 & $6.2 \times 10^{-3}$ \\
    $(10\,000,64)$ & 2.6099 & $10.0 \times 10^{-3}$ \\
    \hline
  \end{tabular}
\end{table}

\section{Conclusion}\label{concludesec}
We derived the equality and bounds for the ASPL of graphs of diameter 3 by using the number of triangles, squares, and some other structures in a graph.
By numerical experiments, we confirmed that the upper and lower bounds obtained are also accurate approximation for the ASPL\@.
On the basis of one of our bounds,
we propose to use $3\triangle + 2\Box$ as the evaluation function of SLS for the ASPL minimization problem where $n$ is at least and close to $d^2$, so that
the random graph has diameter 3 with high probability.
We show that the proposed SLS algorithms requires $O(1)$ time for the evaluation and $O(d^2)$ time for the update.
We construct low ASPL graphs of diameter 3 by IFI and SA using our evaluation function.

\par While the upper bound evaluated by the proposed algorithm is represented only by the number of triangles and squares, the other bounds represented by the number of more complicated structures such as $\triangle^{(2)},\Box^{(2)}$ are confirmed to be more accurate. Thus, if there exists $O(1)$-time evaluation algorithm that calculates one of these bounds, then we can use the bound as the evaluation function of SLS\@.
It would construct graphs with lower ASPL\@.

Finally, we note that our bounds of ASPL for graphs of diameter 3 can be generalized to larger diameter in a similar way.
The construction of low ASPL graphs of larger diameter by similar algorithm is an interesting future work.

\section*{Acknowledgment}
This work was supported by MEXT KAKENHI Grant Number 24106008.

\bibliographystyle{IEEEtran}
\bibliography{IEEEabrv,ref}
\end{document}